\documentclass[epj,final]{svjour}
\usepackage{latexsym,amsmath}
\usepackage{graphics}
\usepackage{graphicx}
\usepackage{amssymb}
\usepackage[latin1]{inputenc}

\begin{document}

\title{Near-$T_c$ second-harmonic emission in high-density bulk MgB$_2$ at microwave frequency}

\author{A. Agliolo~Gallitto\inst{1}\and G. Bonsignore\inst{1}\and G. Giunchi\inst{2}
\and{M. Li~Vigni\inst{1}\institute{CNISM and Dipartimento di
Scienze Fisiche e Astronomiche, Universit$\mathrm{\grave{a}}$ di
Palermo, Via Archirafi 36, I-90123 Palermo (Italy) \and EDISON
S.p.A., Divisione Ricerca \& Sviluppo, Via U. Bassi 2, I-20159
Milano (Italy)}
}}%

\abstract{We discuss the microwave second-harmonic generation in
high-density bulk MgB$_2$, prepared by the reactive liquid Mg
infiltration technology. The intensity of the harmonic signal has
been investigated as a function of temperature and amplitudes of
the DC and microwave magnetic fields. The results are discussed in
the framework of a phenomenological theory, based on the two-fluid
model, which assumes that both the microwave and static magnetic
fields, penetrating in the surface layers of the sample, weakly
perturb the partial concentrations of the normal and
superconducting fluids. We show that, in order to account for the
experimental results, it is essential to suppose that in MgB$_2$
the densities of the normal and condensed fluids linearly depend
on the temperature.
}
\maketitle

\section{Introduction}
Investigation of the nonlinear microwave response of
superconductors is of great interest for both fundamental and
applicative aspects. From the basic point of view, it allows
determining the electron-phonon scattering time \cite{Amato,Ora},
the temperature dependence of the condensed and normal fluids
\cite{noiPRB,PhysC305} and checking the presence of weak links in
the sample \cite{PhysC161,BKBO,jeffries,tinkham}. From the
technological point of view, it is well known that the occurrence
of nonlinear effects at high input-power levels limits the
application of superconductors in passive microwave (mw) devices,
while it has important implications in active devices
\cite{libro1,libro2}. For these reasons, it is of importance
estimating the mw current and field intensities at which nonlinear
effects are significant and recognizing the mechanisms responsible
for the nonlinearity at high frequencies.

Since the discovery of superconductivity in magnesium diboride,
the scientific world has devoted large attention to this
superconductor because of its peculiar properties, related to the
two-gap structure of the electronic states, as well as its
potential in technological applications \cite{bugo,Hein,Coll}. It
has been shown that the presence of the two gaps affects several
properties of the MgB$_2$ superconductor, such as the fluxon
structure \cite{esk,golubov} and the temperature dependencies of
the specific heat \cite{bouquet}, critical fields
\cite{critfield1,critfield2}, condensed-fluid density
\cite{golubov2,Moca}. The advantages of using MgB$_2$ for
technological applications are the simple crystal structure, the
malleability and ductility, due to the metallic nature, the
relatively high critical temperature, the large coherence length,
which makes the materials less susceptible to structural defects
like grain boundaries \cite{bugo,nature2,Khare}. One of the main
effects of the nonlinear electromagnetic response of
superconductors is the emission of signals at harmonic frequencies
of the driving field \cite{samoilova}. Previous studies carried
out in some bulk MgB$_2$ samples have shown that the microwave
harmonic emission is significant in the whole range of
temperatures below $T_c$ and exhibits an enhanced peak at
temperatures close to $T_c$ \cite{PhysC432,noiTH}. It has been
shown that the harmonic emission at low temperatures is related to
processes occurring in weak links, while a different mechanism is
responsible for the near-$T_c$ peak \cite{PhysC432}. Furthermore,
measurements of second-harmonic (SH) signal at DC magnetic fields
smaller than the lower critical field, have shown that the SH
signal due to weak links strongly depends on the specific
properties of the sample \cite{PhysC432,eucas2005}.

Recently, we have investigated the mw SH emission in three bulk
samples of MgB$_2$ obtained by reactive infiltration of liquid Mg
in powdered B preforms \cite{giunchi}, which are characterized by
different mean size of the grains (100, 40, 1 $\mu$m). We have
shown that the intensity of the low-field and low-$T$ SH signal
strongly depends on the size of grains and, in particular, that
the sample with the smallest grain size does not exhibit
detectable low-$T$ SH signal, proving that in such sample the
nonlinear processes in weak links are not enough effective in the
mw SH generation \cite{eucas2005}. In this paper, we investigate
the SH emission at temperatures close to $T_c$ in the sample with
grain mean size of $\approx 1~\mu$m. The investigation is carried
out at low magnetic fields, where nonlinear effects arising from
motion of Abrikosov fluxons do not play a significant role in the
harmonic emission. The results are discussed in the framework of a
phenomenological theory, based on the two-fluid model, which
assumes that both the mw and static magnetic fields, which
penetrate in the surface layers of the sample, weakly perturb the
partial concentrations of the normal and superconducting fluids
\cite{noiPRB,PhysC305}. We show that the results expected from
this model satisfactorily account for the experimental data if a
linear temperature dependence of the normal and condensed fluid
densities is supposed.
\section{Experimental Apparatus and Sample}
The investigated MgB$_2$ sample was prepared by reactive liquid Mg
infiltration technology \cite{giunchi}. Micron size amorphous
boron powder (Grade I, 98\% purity, Stark H.C., Germany) and pure
liquid magnesium are inserted in a stainless steel container, with
a weight ratio Mg/B over the stoichiometric value ($\approx
0.55$); the container was sealed by conventional tungsten inert
gas welding procedure, with some air trapped inside the B powder;
a thermal annealing at 900~°C for 30 min was performed. It has
been shown that MgB$_2$ samples produced in this way have a
density of 2.40~g/cm$^3$ and consists predominantly of 1-micron
size grains, although larger grains, of a few microns in size, are
also present \cite{giunchiSST}. From the final product, we have
extracted a sample of approximate dimensions
$2\times3\times0.5~\mathrm{mm}^3$, whose largest faces have been
mechanically polished to obtain very smooth surfaces.

The sample has been previously characterized by measuring the AC
susceptibility at 100~kHz. From these measurements we have found
that the transition can be described by a Gaussian distribution
function of $T_c$, centered at $T_{c0}=38.5~$K with
$\sigma_{T_c}=~0.2$~K. Furthermore, we have determined the lower
and upper critical fields; in the range of temperatures of about 5
K below $T_c$, the critical fields show linear temperature
dependence with $dH_{c2}/dT \approx 1.5$~kOe/K and
$dH_{c1}/dT\approx 6.5$~Oe/K.

In order to detect the SH signal, the sample is placed in a
bimodal cavity, resonating at the two angular frequencies $\omega$
and $2\omega$, with $\omega/2\pi \approx 3~$~GHz, in a region in
which the mw magnetic fields $\textit{\textbf{H}}(\omega)$ and
$\textit{\textbf{H}}(2\omega)$ are maximal and parallel to each
other. The $\omega$-mode of the cavity is fed by a train of mw
pulses, with pulse width $5~\mu$s, pulse repetition rate 200 Hz,
maximum peak power $\sim 50$~W. The SH signal radiated by the
sample is detected by a superheterodyne receiver. The cavity is
placed between the poles of an electromagnet that generates DC
magnetic fields up to $\approx 10$~kOe; two additional coils,
externally fed, allow reducing the residual field within 0.1 Oe
and working at low magnetic fields. All the measurements have been
performed with
$\textit{\textbf{H}}_0||\textit{\textbf{H}}(\omega)||\textit{\textbf{H}}(2\omega)$.
The experimental apparatus is described in more details in
Ref.~\cite{metamat}.

\section{Experimental Results}
The SH emission has been investigated as a function of the
temperature, DC magnetic field and input power level. In order to
disregard SH signals due to nonlinear fluxon dynamics, the
attention has been devoted to the SH response at low external
fields. Before any measurement was performed, the sample was
zero-field cooled (ZFC) down to a desired temperature value and
then the magnetic field was applied.

Fig. 1 shows the SH signal intensity as a function of the
temperature in the MgB$_2$ sample, at different values of the DC
magnetic field. The SH emission is visible, with our experimental
sensibility, only in a restrict range of temperatures below $T_c$
(the dashed line indicates the noise level). The
SH-\textit{vs}.-\textit{T} curves exhibit a peak just below $T_c$;
on increasing $H_0$, the peak broadens and its maximum shifts
towards lower temperatures.
\begin{figure}[h]
\centering
\includegraphics[width=8cm]{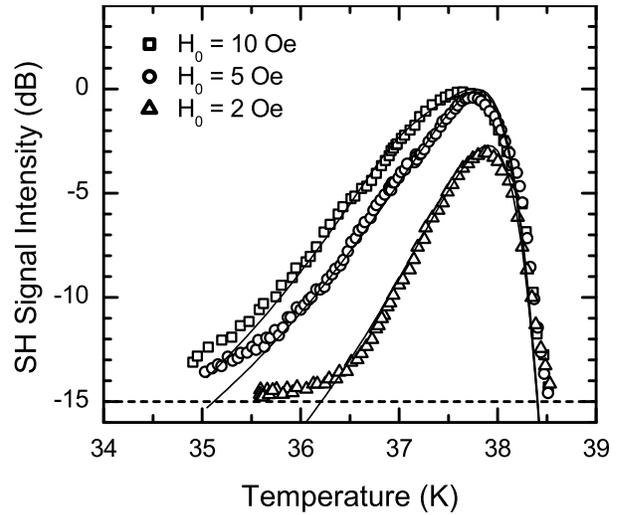}
\caption{SH signal intensity as a function of the temperature in
the MgB$_2$ sample, for $H_0=10$,~5,~2 Oe as displayed from the
top. Input peak power $\approx30$~dBm; the dashed line indicates
the noise level that corresponds to $\approx-75$~dBm. The curves
are normalized to the maximal intensity of the SH signal detected
at $H_0=10$~Oe. The continuous lines have been obtained with the
model discussed in the next section, using
$\lambda_0/\delta_0=0.11,~\alpha_0=2\times10^{-3}
\mathrm{Oe}^{-1},~\alpha_1 H_1 =8\times10^{-3}$.}
\end{figure}

Fig. 2 shows the SH signal intensity as a function of the DC
magnetic field, at the two temperatures $T=37.6$~K and $T=36.7$~K.
The results have been obtained in the ZFC sample on increasing
$H_0$ for the first time. As expected from symmetry
considerations, the SH signal is zero at $H_0=0$; on increasing
the field it rapidly increases, exhibiting a wide maximum at
fields of few Oe. Both the slope of the SH-\textit{vs}.-$H_0$
curves at $H_0=0$ and the position of the maximum depend on the
temperature.
\begin{figure}[tbh]
\centering
\includegraphics[width=8cm]{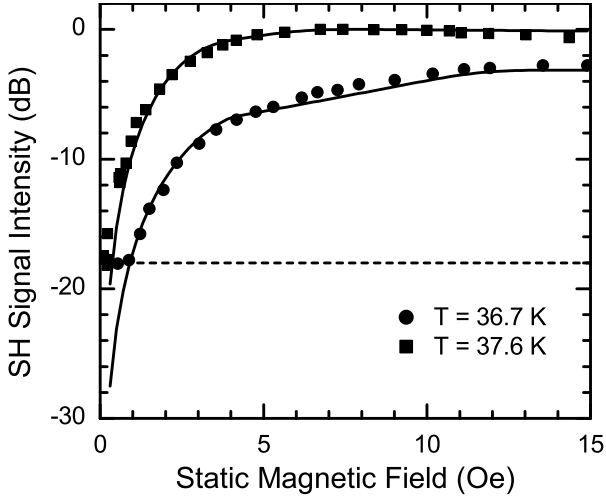}
\caption{SH signal intensity as a function of the static magnetic
field, for $T = 37.6$~K and $T = 36.7$~K, as displayed from the
top. Input peak power $\approx$~30~dBm. The curves are normalized
to the maximal intensity of the SH signal detected at $T =
37.6$~K. The dashed line indicates the noise level. The continuous
lines have been obtained by the model discussed in the next
section, using the same values of the parameters of Fig.~1.}
\end{figure}

By sweeping the magnetic field from zero up to a certain value,
$H_{max}$, and back, the SH signal shows a hysteretic behavior
when $H_{max}$ reaches a threshold value that depends on
temperature. Fig. 3 shows the field dependence of the signal
obtained sweeping $H_0$ in the range $\pm 15$~Oe, for $T=37.6$~K .
The presence of a magnetic hysteresis, as well as the observation
of a noticeable SH signal at $H_0=0$, in the decreasing-field
branch of the curve, suggest that trapped flux is present in the
sample. So, we deduce that, even at temperatures near $T_c$, a
critical state of the fluxon lattice develops. It is worth to
remark that, on further increasing the DC magnetic field, the
intensity of the SH signal decreases and, eventually, goes to zero
when $H_0$ reaches the value of the upper critical field.

\begin{figure}[b]
\centering
\includegraphics[width=8cm]{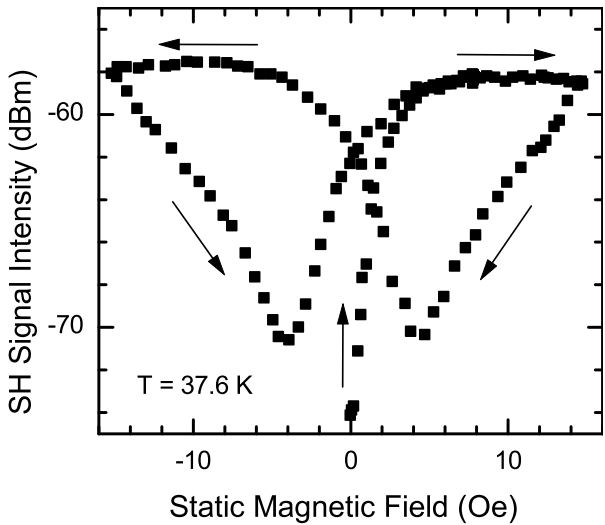}
\caption{SH signal intensity as a function of the static magnetic
field, obtained sweeping $H_0$ in the range $\pm 15$~Oe. Input
peak power $\approx$~30~dBm.}\label{farfalla}
\end{figure}

In Fig. 4 we report the power dependence of the SH signal at
$H_0=2$~Oe, for two different values of the temperature. As one
can see, the power dependence of the SH signal cannot be described
by a mere \textit{n}-order power law, indeed the slope of the
SH-\textit{vs}.-$P_{in}$ line is not constant in the range of
power levels investigated; in particular, at the investigated
power levels the input-power dependence of the SH signal is less
than quadratic.
\begin{figure}[thb]
\centering
\includegraphics[width=8cm]{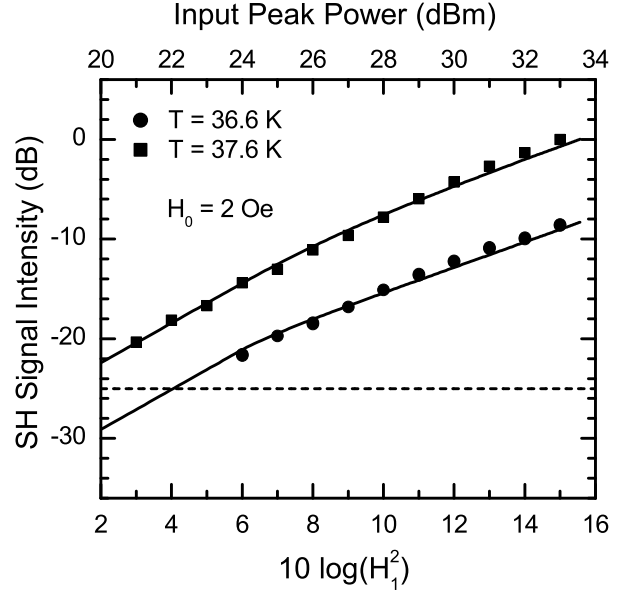}
\caption{SH signal intensity as a function of the input peak power
for $H_0=2$~Oe and two different values of the temperature. The
continuous lines have been obtained by the model discussed in the
next section, using $\lambda_0/\delta_0=0.11,~
\alpha_1=\alpha_0=2\times10^{-3} \mathrm{Oe}^{-1}$. The dashed
line indicates the noise level.}
\end{figure}

\section{The Model}
The nonlinear electromagnetic response has been investigated in
conventional as well as in high-$T_c$ superconductors
\cite{samoilova}. Several mechanisms responsible for the harmonic
emission have been recognized; their effectiveness depends on the
temperature, DC magnetic field and type of superconductor. At DC
fields smaller than the lower critical field, the harmonic
emission at $T \ll T_c$ has been ascribed to nonlinear processes
occurring in weak links \cite{PhysC161,BKBO,jeffries}, while at
temperatures close to $T_c$ it has been ascribed to modulation of
the order parameter induced by the mw magnetic field
\cite{Amato,noiPRB,PhysC305,BKBO,truninfrance}.

Recently, we have investigated bulk MgB$_2$ samples prepared by
different methods and in all of them we have detected a near-$T_c$
peak in the SH emission; however, in most of them the peak mingles
with the low-\textit{T} signal \cite{PhysC432}. As it has already
been mentioned, the sample here investigated does not exhibit
detectable SH emission at low fields and low temperatures, showing
that the grain boundaries are not enough effective to give rise SH
emission related to processes occurring in weak links. This
finding ensures the high sample quality and allows investigating
the mechanism responsible for the harmonic emission near $T_c$ in
the MgB$_2$ superconductor.

Nonlinear emission at temperatures close to $T_c$ is expected in
the framework of several models, some work in the weak-coupling
limit \cite{Ora,GE27,GE28}, others in the strong-coupling limit
\cite{truninfrance}, others in the framework of the two-fluid
model \cite{noiPRB,PhysC305,BKBO}. In all of them, the harmonic
emission originates from the time variation of the order parameter
induced by the em field.

From the models reported in
Refs.~\cite{Ora,truninfrance,GE27,GE28}, the input power
dependence of the \textit{n}th-harmonic signal is expected to
follow an \textit{n}-order power law. As it is shown in Fig. 3, we
have observed a less than quadratic power dependence of the SH
signal intensity; furthermore, the results reported in
Ref.~\cite{noiTH} show a less than cubic power dependence of the
near-$T_c$ third-harmonic signal in MgB$_2$. So, these models
cannot justify the near-$T_c$ harmonic emission in MgB$_2$.

Results of second and third harmonic emission near $T_c$ by
high-$T_c$ superconductors have been justified quite well in the
framework of the two-fluid model, with the additional hypothesis
that both the mw and the DC magnetic fields, penetrating the
sample in the surface layers, linearly perturb the partial
concentrations of the normal and condensed fluids \cite{noiPRB}.
Since the processes occur in the surface layer of the order of the
field penetration depth, they are effective only at temperatures
close to $T_c$, where the energy gap is not large and the
penetration depth is of maximal extent. The model has been
discussed for the first time by Ciccarello \textit{et al.}
\cite{noiPRB} to account for the experimental data of
third-harmonic emission by YBa$_2$Cu$_3$O$_7$ (YBCO) single
crystals in the absence of DC magnetic fields. Successively, it
has been generalized to take into account the effect of the DC
magnetic field and, further, justify the SH generation
\cite{PhysC305,BKBO}. The model is valid in the limit of strong
coupling, in which the approach of the two-fluid model can be
reasonably used \cite {35BKBO}. It has been shown that the model
accounts very well for the results obtained in high-$T_c$
superconductors in the Meissner state and qualitatively those
obtained in the mixed state \cite{PhysC305}. So, although it has
been shown that the mechanism contributes to the harmonic emission
also in the mixed state, others processes, related to the presence
and motion of fluxons, may contribute to the nonlinear mw response
of superconductors in the mixed state \cite{CCSH,PhysC159}.

Since the near-$T_c$ SH signal radiated by MgB$_2$ shows some
peculiarities similar to those observed in YBCO and
Ba$_{0.6}$K$_{0.4}$BiO$_3$ (BKBO) crystals \cite{PhysC305,BKBO},
such as the input power dependence and the broadening of the peak
on increasing $H_0$, one can reasonably hypothesize that in these
superconductors the mechanism responsible for the near-$T_c$ SH
emission is the same. However, few differences have to be
mentioned: the SH emission is more enhanced in YBCO than in
MgB$_2$ (at least one order of magnitude); the near-$T_c$ peak is
narrower in YBCO ($\sim T_c/30$) than in MgB$_2$ ($\sim T_c/10$);
while in YBCO and BKBO the peak width, as well as the position of
the maximum, depend on the input power, these effects have not
been detected in MgB$_2$. In the following of this section, we
describe the model.

Coffey and Clem \cite{CC} have elaborated a comprehensive theory
for the electromagnetic response of type-II superconductors in the
framework of the two-fluid model of the superconductivity. It has
been shown that the em field is characterized by a complex
penetration depth
\begin{equation}\label{lamtil}
\widetilde{\lambda}^2=\frac{\lambda^2+\delta_{v}^2}
{1-2i\lambda^2/\delta^2},
\end{equation}
where

\begin{equation}\label{lamda}
\lambda = \frac{\lambda_0}{\sqrt{(1-w_0)(1- B_0 /H_{c2})}},
\end{equation}
\begin{equation}\label{delta}
\delta = \frac{\delta_0}{\sqrt{1-(1-w_0)(1- B_0 /H_{c2})}}.
\end{equation}
In Eqs.~(2) and (3), $\lambda_0$ is the London penetration depth
at $T=0$; $\delta_0=(c^2/2 \pi \omega \sigma_0)^{1/2}$ is the
normal skin depth at $T = T_c$; $w_0$ and $1- w_0$ are the
fractions of normal and superconducting electrons at $H_0 = 0$, in
the Gorter and Casimir two-fluid model $w_0=(T/T_c)^4$. In
Eq.~(1), $\delta_v$ is the effective complex skin depth arising
from the vortex motion. At mw frequencies and at temperatures
close to $T_c$ it is reasonable to assume that vortices move in
the flux-flow regime; in this case, using the expression of the
viscous-drag coefficient given by Bardeen and Stephen
\cite{bardinS}, it results $\delta_v^2=2\delta_0 B_0/B_{c2}(T)$,
where $B_0$ is the magnetic induction due to the presence of
fluxons.

For a sample of thickness $D$ much greater than the characteristic
penetration depths, the mw magnetic induction averaged over the
volume of the crystal is
\begin{equation}\label{B}
<B(t)> = \frac{-H_1}{D}[\Im({\widetilde{\lambda}})\cos{\omega
t}+\Re(\widetilde{\lambda}) \sin{\omega t}],
\end{equation}
where $H_1$ is the amplitude of the mw field. \\ The Fourier
coefficients of $<B(t)>$ are given by
\begin{equation}\label{an}
a_n = \frac{1}{\pi} \int_0^{2\pi} <B(t)> \cos (n \omega t)
d(\omega t),
\end{equation}
\begin{equation}\label{bn}
b_n = \frac{1}{\pi} \int_0^{2\pi} <B(t)> \sin (n \omega t)
d(\omega t).
\end{equation}
From $a_n$ and $b_n$ one can calculate the induced magnetization
oscillating at $n\omega$ and, eventually, the intensity of the
\textit{n}th-harmonic signal.

The magnetic induction obtained using expressions (1)-(3) does not
contain harmonic Fourier components because all the equations
involved up to now are linear. Nonlinearity comes out on assuming
that the mw field, penetrating the surface layers of the sample,
modulates the partial concentrations of the normal and
superconducting fluids \cite{noiPRB}.

In the Meissner state, also the DC magnetic field penetrates in
the surface layers of the sample and it can perturb the partial
concentration of the two fluids. On the other hand, though we have
performed measurements at low magnetic fields ($H_0 \sim 10$ Oe),
at temperatures very close to $T_c$ the sample goes in the mixed
state; in this case, a part of the DC magnetic field penetrates as
fluxons and the other is screened by the sample and is present in
the surface layers. We suppose that both the mw and the screened
DC magnetic field perturb the electron concentrations in a similar
way. Following this idea, it has been set \cite {PhysC305}
\begin{equation}\label{w(t)}
w(t, H_0)= w_0[1+|\alpha_0 4\pi M + \alpha_1 H_1 \cos{\omega t}|],
\end{equation}
where $\alpha_0$ and $\alpha_1$ are phenomenological parameters.
\\
By replacing $w_0$ in Eqs.~(\ref{lamda}) and (\ref{delta}) with
$w(t, H_0)$ of Eq.~(\ref{w(t)}), the em induction inside the
sample is expected to have harmonic Fourier components, which can
be calculated by using Eqs.~(\ref{an}) and (\ref{bn}). The power
emitted by the sample at $n\omega$ is expected to be:
\begin{equation}\label{Pn}
P_{n \omega}\propto a_n^2 + b_n^2.
\end{equation}

\section{Discussion}
By using Eqs.~(\ref{B}-\ref{Pn}) we have deduced the expected SH
signal intensity. The integrals of Eqs.~(\ref{an}) and (\ref{bn})
cannot be solved analytically; they have been solved numerically
with the proviso that $w(t, H_0)=1$ whenever, because of the
contribution of the perturbative terms in Eq.~(\ref{w(t)}), its
instantaneous value becomes greater than one. This condition is
necessary since the concentration of normal electrons cannot
exceed one and no modulation must occur when all electrons are in
the normal state.

The calculations have been carried out in the following
approximation: $|4 \pi M|$ equal to $H_0$ for $H_0\leq H_{c1}(T)$;
$|4 \pi M|$ linearly decreasing from $H_{c1}(T)$ to zero for
$H_{c1}(T)\leq H_0\leq H_{c2}(T)$.

The expected results depend on the ratio $\lambda_0/\delta_0$, the
temperature dependence of $w_0$ and the parameters $\alpha_0$ and
$\alpha_1$. In particular, $\lambda_0/\delta_0$ and the
temperature dependence of $w_0$ determine the peak width,
$\alpha_0$ and $\alpha_1$ the field and power dependence of the SH
signal, respectively. Since the width of the near-$T_c$ peak is
most likely affected by the inhomogeneous broadening of the
superconducting transition, we have calculated the averaged em
induction (Eq.~\ref{B}) taking into account the distribution of
$T_c$ over the sample. By using reasonable values of
$\lambda_0/\delta_0$ and the temperature dependence of $w_0$
expected from the Gorter and Casimir two-fluid model, we have
obtained a near-$T_c$ peak much narrower than the one
experimentally observed. On the other hand, different authors
\cite{golubov2,Moca,ecc,ecc2} have shown that the temperature
dependence of the field penetration depth in MgB$_2$ cannot be
accounted for by either the Gorter and Casimir two-fluid model or
the standard BCS theory. A linear temperature dependence of the
condensed fluid density, in a wide range of temperature below
$T_c$, has been reported, which has been justified in the
framework of two-gap models for the MgB$_2$ superconductor
\cite{golubov2,Moca}.

Prompted by these considerations, we have calculated the averaged
em induction over the sample, from Eq.~(\ref{B}), supposing a
linear temperature dependence of $w_0$ and using
$\lambda_0/\delta_0$, $\alpha_0$ and $\alpha_1$ as parameters. The
continuous lines of Figs.~1 and 2 show the temperature and field
dependence of the SH signal, respectively, expected from the
model. The best-fit curves have been obtained with
$\lambda_0/\delta_0=0.11,~\alpha_0=2\times 10^{-3}$~ Oe$^{-1}$ and
$\alpha_1 H_1=8\times 10^{-3}$. Since the amplitude of the mw
magnetic field at which we have performed the measurements is
about 4~Oe, we may estimate $\alpha_1$ to be of the same order of
$\alpha_0$.  By using the same value of $\lambda_0/\delta_0$ and
setting $\alpha_1=\alpha_0=2\times10^{-3}$~Oe$^{-1}$ we have
obtained the continuous lines of Fig.~4 for the power dependence
of the SH signal intensity. As one can see, the results expected
from the model, hypothesizing that the normal and condensed fluid
densities linearly depend on temperature, satisfactorily agree
with the experimental data. We would remark that, by using the
same values of the parameters before mentioned, we have found
that, contrary to what occurs in other superconductors
\cite{noiPRB,PhysC305,BKBO}, the width of the SH near-$T_c$ peak
in MgB$_2$ is not appreciably affected by the amplitude of the
microwave magnetic field. This finding, which agrees with the
experimental observations, may be ascribed to the different values
of the parameters used to fit the experimental data of MgB$_2$,
with respect to those obtained in other superconductors.

Since $\lambda_0/\delta_0= \sqrt{\omega\tau/2}$, by supposing
$\lambda_0$ of the order of 100~nm we may estimate the scattering
time of the normal electron to be $\tau\sim 10^{-12}$~s, which is
of the same order of that reported in the literature for MgB$_2$
\cite{tau}. However, we would remark that we have assumed a linear
temperature dependence of the condensed fluid density, which could
be no longer valid at low temperatures; so, it is possible that
the value of $\lambda_0/\delta_0$ we found is not exactly equal to
$\sqrt{\omega\tau/2}$.

The hysteretic behavior shown in Fig.~\ref{farfalla} suggests that
mechanisms related to trapped magnetic flux are effective. On the
other hand, it is reasonable to hypothesize that at temperatures
near $T_c$ the sample goes in the mixed state for magnetic fields
of the order of 10~Oe. The presence of the magnetic hysteresis
suggests that a critical state of the fluxon lattice develops in
the sample up to temperatures near $T_c$. Since the model here
reported does not consider such effects, we did not fit these
results. When the sample is in the critical state, the magnetic
induction, $B_0$, is different for applied magnetic field reached
at increasing and decreasing field, causing a magnetic hysteresis
in the SH signal. The sharp minima observed after the inversion of
the field-sweep direction may be ascribed to the change of the
critical current in the sample skin layer \cite{truninfrance}.

\section{Conclusions}
In conclusion, we have reported a set of experimental results on
second harmonic generation at temperatures close to $T_c$ in a
bulk sample of MgB$_2$, prepared by the reactive liquid Mg
infiltration technology. The SH signal has been investigated as a
function of temperature and amplitudes of the DC and mw magnetic
fields. The results have been discussed in the framework of a
phenomenological model previously elaborated to account for the
nonlinear emission near $T_c$ by YBCO and BKBO crystals. The model
assumes that at temperatures close to $T_c$ the mw and the DC
fields, penetrating in the surface layers of the sample, perturb
the partial concentrations of the normal and superconducting
electrons. We have shown that, in order to account for the
experimental results, it is essential to suppose that in MgB$_2$
the densities of the normal and condensed fluids linearly depend
on the temperature. Such finding agrees with experimental results
of the temperature dependence of the penetration depth reported in
the literature, which have been justified in the framework of
two-gap models for MgB$_2$ superconductor. We have shown that the
enhanced nonlinear effects detected near $T_c$ in MgB$_2$ have the
same origin of those observed in YBCO and BKBO crystals. Though
the superconductors have different nature, the description of the
electromagnetic properties in terms of the two-fluid model seams
to be appropriate. The different peculiarities of the near-$T_c$
peak in the SH response, among the investigated superconductors,
have to be ascribed to a different temperature dependence of the
normal and condensed fluid densities.

\section*{Acknowledgements} The authors are very glad to thank E. Di Gennaro
for his interest and helpful suggestions; G. Lapis and G. Napoli
for technical assistance.

\begin{flushright}\today \end{flushright}

\end{document}